\documentclass{WileyMSP-template}

\usepackage{graphicx}%
\usepackage{multirow}%
\usepackage{amsmath,amssymb,amsfonts}%
\usepackage{amsthm}%
\usepackage{mathrsfs}%
\usepackage[title]{appendix}%
\usepackage{xcolor}%
\usepackage{textcomp}%
\usepackage{manyfoot}%
\usepackage{booktabs}%
\usepackage{algorithm}%
\usepackage{algorithmicx}%
\usepackage{algpseudocode}%
\usepackage{listings}%
\usepackage{siunitx}
\DeclareSIUnit\angstrom{\text{\AA}}
\usepackage{adjustbox}
\usepackage{caption}
\usepackage{hyperref}
\usepackage[capitalise,noabbrev]{cleveref}
\crefname{Figure}{Figure}{Figure}
\crefname{Equation}{Equation}{Equation}
\usepackage{float}
\usepackage{lmodern}
\usepackage{anyfontsize}
\usepackage{comment}

\usepackage[square, super, numbers, sort&compress]{natbib}

\newcommand{\scolor}{black}
\newcommand{\revision}{\textcolor{black}}

\begin{document}

\pagestyle{fancy}
\rhead{\includegraphics[width=2.5cm]{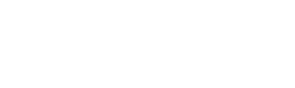}}

\title{BInD: Bond and Interaction-generating Diffusion Model for Multi-objective Structure-based Drug Design}

\maketitle

\author{Joongwon Lee\textsuperscript{$\dagger$}} 
\author{Wonho Zhung\textsuperscript{$\dagger$}}
\author{Jisu Seo}
\author{Woo Youn Kim*}

\begin{affiliations}
J. Lee, W. Zhung, J. Seo\\
Department of Chemistry, KAIST, Yuseong-gu, 34141, Daejeon, Republic of Korea\\

Prof. W. Y. Kim\\
Department of Chemistry, KAIST, Yuseong-gu, 34141, Daejeon, Republic of Korea\\
Graduate School of Data Science, KAIST, Yuseong-gu, 34141, Daejeon, Republic of Korea\\
HITS Inc., 28 Teheran-ro 4-gil, Gangnam-gu, 06234, Seoul, Republic of Korea\\
Email: wooyoun@kaist.ac.kr
\end{affiliations}

\dedication{$\dagger$ These authors contributed equally to this work.}


\begin{abstract}
Recent remarkable advancements in geometric deep generative models, coupled with accumulated structural data, enable structure-based drug design (SBDD) using only target protein information. However, existing models often struggle to balance multiple objectives, excelling only in specific tasks. BInD, a diffusion model with knowledge-based guidance, is introduced to address this limitation by co-generating molecules and their interactions with a target protein. This approach ensures balanced consideration of key objectives, including target-specific interactions, molecular properties, and local geometry. Comprehensive evaluations demonstrate that BInD achieves robust performance across all objectives, matching or surpassing state-of-the-art methods. Additionally, we propose an NCI-driven molecule design and optimization method, enabling the enhancement of target binding and specificity by elaborating the adequate interaction patterns.
\end{abstract}

\section{Introduction}\label{sec1}

Recent advances in geometric deep generative models \cite{isert2023structure, dgl_sbdd_review} on top of the accumulation of biological structures \cite{berman2008protein, burley2019rcsb} have fueled a new paradigm of deep learning-driven structure-based drug design (SBDD), which take advantages of the rich context of protein structures to design well-binding molecules with their 3D poses. 
\revision{In comparison to the earlier molecular generative models that design compounds by producing a string representation of molecular graphs, \cite{dldd_review,ligdream} incorporating the surrounding protein environment as an inductive bias markedly enhances the quality of generation}.
\cite{dgl_sbdd_review} Traditional approaches must rely on separate conformer generation and docking tools to validate whether they can fit in a pocket.
In contrast, designing a molecule directly in a protein pocket is a strong constraint that inherently enforces the binding of the designed molecule, or at least ensures that its 3D shape is compatible with the pocket, thereby paving the way for more reliable SBDD.

\begin{figure}[!ht]
    \centering
    \includegraphics[width=0.95\linewidth]{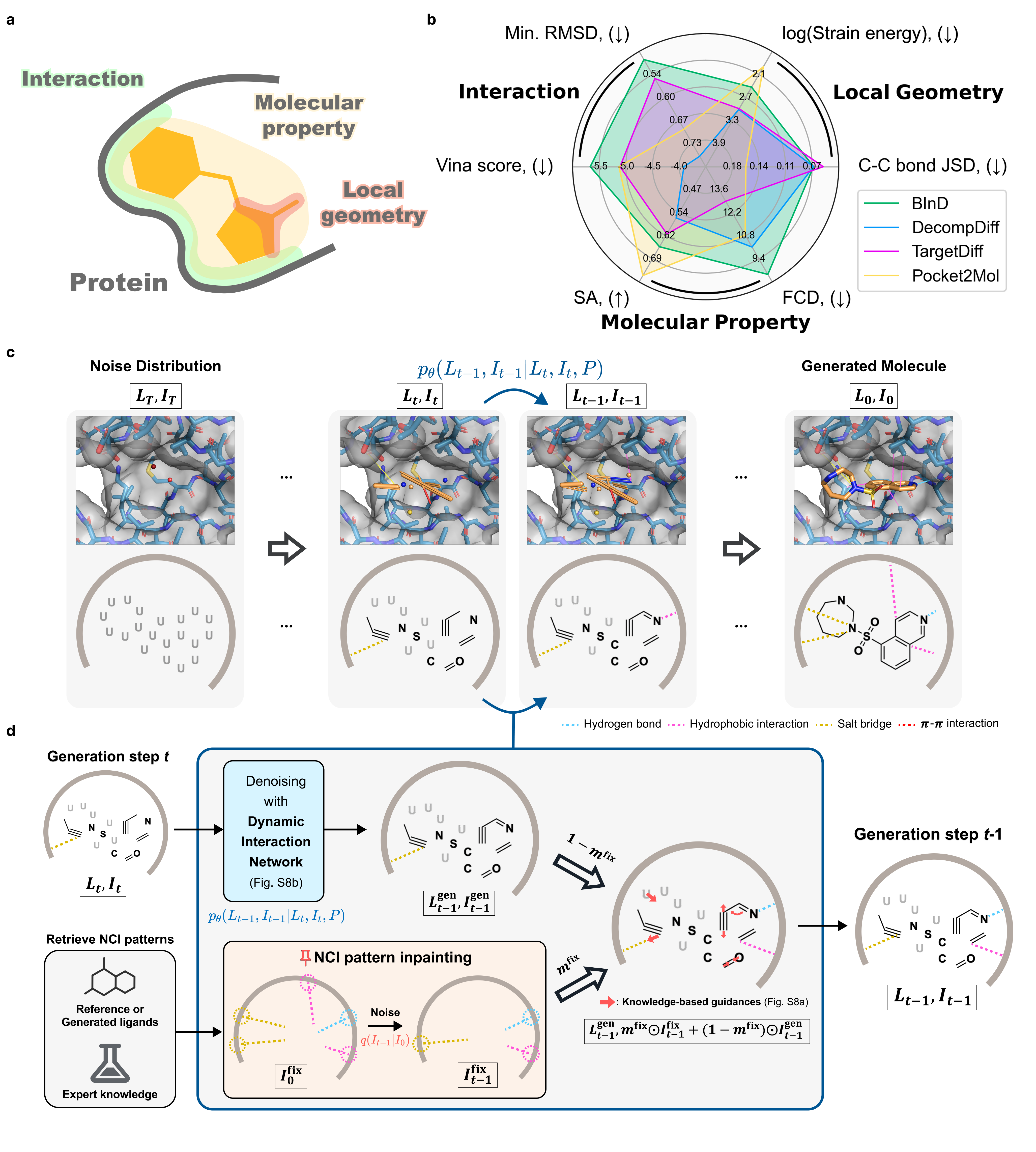}
    \caption{\textbf{
    Overview of BInD.
    } 
    \textbf{a}, Conceptual illustration of the three key objectives of deep SBDD: accurate local geometry, desirable molecular property, and target-specific interactions.
    \textbf{b}, Overall performance of deep SBDD models. While other baseline models fall short in at least one of the three key objectives, BInD shows balanced performance with notable strength in considering interactions. 
    \textbf{c}, An overview of the generative process of BInD, where atoms, bonds, and interactions are denoised explicitly and simultaneously. The 2D illustration below focuses on the types of each entity, where dashed lines indicate NCIs and `U' indicates an absorbing type.
    \textbf{d}, A detailed illustration of a single generation step, in case the desired interaction pattern is given. A generating molecule at $t$ step is first denoised with a dynamic interaction network ($p_\theta$), then integrated with the given NCIs noised with a forward diffusion ($q)$. Then, knowledge-based guidance terms are applied to modify atom positions finely, obtaining a molecule at $t-1$ step. 
    }
    \label{fig:overview}
\end{figure}

However, existing SBDD models often struggle to meet the \revision{multiple} requirements to be a drug candidate with a realistic 3D structure. \cite{PoseCheck,molcraft} 
\revision{These requirements are categorized into three main objectives:}
1) accurate local geometry, 2) desirable drug-like molecular properties, and 3) formation of target-specific interactions (\cref{fig:overview}a). 
Local geometry refers to how accurately the model generates a local 3D structure aligned with its molecular graph without violating physical rules.
Unrealistic geometry would possess high strain, affecting the reliability of generated molecules, as illustrated in Figure \textcolor{\scolor}{S1}a. \cite{PMDM, hierdiff, PoseCheck}  
Auto-regressive models that sequentially add atoms have demonstrated their strengths \revision{on local geometries} by focusing on adjacent atoms at each generation step. \cite{AR, Pocket2Mol, graphBP}
Molecular properties, in contrast, depend entirely on the \revision{global} structure of a molecular graph. For instance, drug-likeness cannot be determined by examining just part of a molecule; it requires consideration of the entire structure. \cite{QED} 
\revision{Bond information plays a key role in bridging the gap between the generated molecular graph and its 3D conformation. Point cloud-generating approaches often struggle to align the obtained molecular graphs and their 3D pose when relying on post-bond order assigning tools, because even a small error in atom positions can result in a different molecule with undesirable properties.} 
Notably, previous bond co-generating approaches that generate a 3D molecular graph in a one-shot behavior improve both local geometry and \revision{global} molecular properties. \cite{Pocket2Mol, MolDiff, DecompDiff, shepherd}

\revision{A \textit{balanced generative model} refers to a generative framework that simultaneously optimizes multiple interdependent objectives without disproportionately favoring one at the expense of others. In the context of SBDD, achieving such a balance is critical, as performance trade-offs across these axes can impair the overall quality and reliability of the designed molecule.}
\revision{Previous deep learning-based SBDD approaches are often tuned to excel in one or two aspects, but compromise the other criteria. Furthermore, balancing the objectives of SBDD presents significant challenges due to inherent conflicts between metrics. For instance, synthetic accessibility tends to favor smaller, less complex molecules; conversely, achieving high binding affinity often requires larger, more complex molecules that can form extensive interactions with the target protein. \cite{drughive, pilot}}
\revision{These limitations necessitate novel approaches that can jointly optimize the aforementioned objectives without severely sacrificing individual performance metrics.}

In light of this, we propose BInD (\textbf{B}ond and \textbf{In}teraction-generating \textbf{D}iffusion model) with well-balanced performance across the three key objectives. The size of a chemical space that fulfills all three goals is much smaller than that of which a deep SBDD model may represent without any inductive bias. Thus, a desirable model needs a certain constraint or prior knowledge to help it focus on a specific chemical space by precisely understanding the relationship between 3D molecular structures and the three objectives. Inspired by recent perspectives in the working principle of multi-task learning \cite{contravariance, platonic} and its successful demonstration on 3D molecular generations, \cite{MolDiff, midi, shepherd} we leverage task-driven constraints to narrow down the solution space of generated molecules by incorporating concurrent training on 3D molecular graphs and NCI patterns. We focus on a molecule’s global and local 3D structure, critical for forming favorable interactions with a target protein, while its molecular graph governs its chemical properties. BInD bridges these aspects by using diffusion-based generative modeling to simultaneously denoise bond and NCI types along with atom types and positions with knowledge-based guidance. Our approach ensures alignment between key objectives, resulting in more realistic molecular designs. As a result, BInD outperforms or is comparable to state-of-the-art deep SBDD approaches for each task for different purposes and shows balanced performance for the three main objectives (\cref{fig:overview}b). Furthermore, we propose \revision{a NCI-driven molecule design and optimization} method that retrieves favorable NCI patterns from the generated molecules.
As a proxy for a binding preference, the patternized information of NCIs can elicit an enhancement in binding affinities without further training or external evaluation methods. Finally, we apply our modeling strategy in a practical setting to demonstrate the role of NCIs in rendering a target specificity, suggesting the extendability of our strategy to even more challenging drug design tasks.

Our main contributions are as follows.
\begin{itemize}
    \item We introduce BInD, the first end-to-end diffusion-based generative framework that balances key objectives in deep SBDD by explicitly imposing covalent and non-covalent conditions as knowledge-based guidance to align a 2D molecular graph with its 3D pose in a binding pocket.  
    \item We comprehensively evaluated deep SBDD models on each key objective, elucidating the effect of the knowledge-based guidance on the model performance.
    \item We propose an NCI-driven molecule \revision{design and optimization method} and apply it to a practical setting case to achieve target specificity, highlighting the potential of our framework in designing mutant-selective drugs.
\end{itemize}

\section{Results}\label{sec2}

\subsection{Model overview of BInD}
Our main goal is to tackle multi-objective design problems that deep SBDD models face. In this respect, we propose a diffusion-based molecular generative framework named BInD (\textbf{B}ond and \textbf{In}teraction-generating \textbf{D}iffusion model) that simultaneously generates atoms, bonds, and NCIs relative to a given target pocket (\cref{fig:overview}c). Alongside the concurrent generation of each entity, we leverage strategies to assist the model in aligning the molecular graph with its 3D pose (\cref{fig:overview}d). A dynamic interaction network enables BInD to perceive a molecule in a longer or shorter range, reducing the gap between global and local features. Knowledge-based guidance terms act on the generative process to finely steer atom positions within a range of realistic bonds and NCIs. Together, BInD can achieve well-rounded performance in various tasks, addressing the three key objectives in \cref{fig:overview}b, while other models performed well in their specialized tasks. Details of BInD are provided in the Methods section.

\subsection{A comprehensive evaluation of deep SBDD models on three perspectives}\label{sec:benchmark}
Here, we comprehensively categorize benchmark metrics across three key objectives. By comparing BInD with other baseline models, we elucidate the roles of our strategies in accomplishing each goal.
The baseline models are classified into two groups, depending on whether a model utilizes the information of the reference ligands in the test set. 
\revision{
Although BInD does not rely on a reference ligand, it also offers an inpainting mode, termed BInD\textsuperscript{ref}, which generates new compounds guided by the NCI pattern of a reference ligand.}
\revision{Therefore,} we compare BInD\textsuperscript{ref} with reference ligand-dependent models, such as InterDiff, which \revision{acquires} the NCI patterns of reference ligands as an \revision{input} condition, and DecompDiff\textsuperscript{ref}, which centers prior atoms on the fragments of the reference ligand. 
Details on these baseline models are provided in the Method section.

\begin{table*}[!ht]
    \centering
    \scriptsize 
    
    \begin{adjustbox}{max width=\textwidth}
    
    \begin{tabular}{l|cc|cc|cc|cc|cc}
        
        \toprule
        
        \multirow{2}{*}{Model} & \multicolumn{2}{c|}{Vina Score ($\downarrow$)} & \multicolumn{2}{c|}{Vina Min. ($\downarrow$)} & \multicolumn{2}{c|}{Vina Dock ($\downarrow$)} & \multicolumn{2}{c|}{Higher [\%] ($\uparrow$)} & \multicolumn{2}{c}{Atom Count} \\

        & Avg. & Med. & Avg. & Med. & Avg. & Med. & Avg. & Med. & Avg. & Med. \\

        \midrule

        
        
        AR & \textbf{-5.28} & -5.85 & -5.81 & -6.05 & -6.75 & -6.68 & 37.6 \% & 30.0 \% & 18.1 & 17.0 \\ 
        
        Pocket2Mol & -5.13 & -4.81 & \textbf{-6.56} & -6.15 & -7.43 & -7.16 & 51.3 \% & 55.0 \% & 18.6 & 17.3 \\ 

        DiffSBDD & -2.71 & -4.25 & -4.84 & -5.15 & -6.51 & -6.66 & 39.2 \% & 33.0 \% & 20.4 & 20.6 \\
        
        TargetDiff & -5.11 & -6.02 & -6.43 & -6.69 & \textbf{-7.46} & \textbf{-7.82} & 52.6 \% & 50.8 \% & 22.8 & 23.1 \\

        DecompDiff & -3.91 & \textbf{-6.10} & -6.44 & \textbf{-7.31} & \textbf{-8.51} & \textbf{-8.78} & \textbf{72.2 \%} & \textbf{89.8 \%} & 35.2 & 34.0 \\ 

        \textbf{BInD} & \textbf{-5.64} & \textbf{-6.22} & \textbf{-6.56} & \textbf{-6.80} & \textbf{-7.46} & -7.66 & \textbf{53.6 \%} & \textbf{56.3 \%} & 23.9 & 23.8 \\
        
        \midrule
        
        InterDiff & -2.25 & -4.48 & -4.87 & -5.55 & -7.06 & -7.03 & 39.8 \% & 30.0 \% & 25.1 & 24.0 \\ 

        DecompDiff\textsuperscript{ref} & -5.30 & -5.58 & -6.31 & -6.41 & -7.42 & \textbf{-7.65} & 50.3 \% & 47.2 \% & 24.2 & 23.0 \\ 
        
        \textbf{BInD\textsuperscript{ref}} & \textbf{-5.99} & \textbf{-5.91} & \textbf{-6.65} & \textbf{-6.43} & \textbf{-7.50} & -7.55 & \textbf{54.3 \%} & \textbf{59.0 \%} & 22.8 & 21.5 \\

        \midrule

        Reference & -6.36 & -6.46 & -6.71 & -6.49 & -7.45 & -7.26 & - & - & 22.8 & 21.5 \\
        
        \bottomrule
        
    \end{tabular}
    
    \end{adjustbox}
    
    \caption{
    \textbf{Benchmark result of BInD and baseline models on binding affinities.}
    The first six models are reference-free methods, and the last three are dependent methods. The mean and median values of Vina energies, higher ratio, and atom counts are reported. The top 2 performing models among the reference-free methods and the best-performing model among the reference-dependent models for each benchmark are marked as bold.}
    \label{tab:gen_reeval}
\end{table*}

\begin{figure}[!ht]
    \centering
    \includegraphics[width=0.9\linewidth]{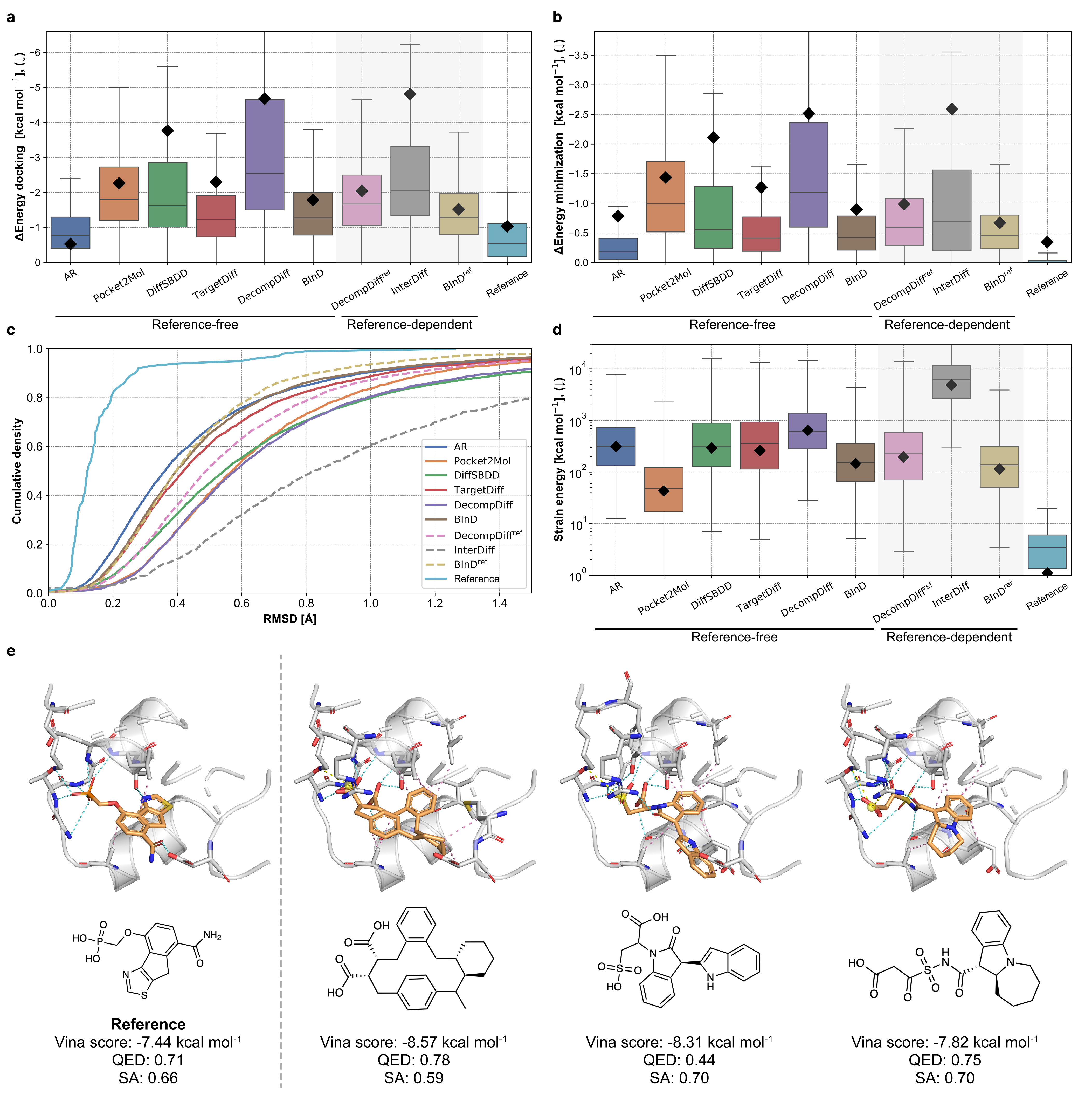}
    \caption{
    \textbf{Comprehensive assessment of BInD with baseline models.}
    \textbf{a}, The box plot illustrates energy differences between before and after Vina minimization. 
    \textbf{b}, The box plot illustrates energy differences between before and after Vina re-docking. 
    \textbf{c}, Cumulative distribution function (CDF) of minimization RMSDs. 
    \textbf{d}, The box plot illustrates strain energies.
    In each box plot, the median and four quartile values are denoted with black diamonds that indicate average values.
    Models are classified into two groups, depending on whether or not they use information from reference ligands. The reference-dependent models are differentiated as a shaded region in box plots and as dashed lines in CDF. \textbf{e}, Three examples of BInD-generated molecules for the test pocket (PDB ID: 3KC1), its NCIs, and its Vina score, QED, and SA. 
    Generated molecules exhibit higher Vina scores with preferable QED and SA scores compared to the reference.
    In terms of NCI, BInD generates molecules with hydrogen bonds targeting hydrophilic regions at the loop between two helices. Also, the molecules form new NCIs that are not present in the reference molecule, such as a salt bridge with an arginine.
    }
    \label{fig:gen_benchmark}
\end{figure}

\noindent\textbf{Interaction with a target protein}. First, we evaluated the binding affinities of generated molecules with various target proteins. 
As in previous studies, \cite{Pocket2Mol, DiffSBDD, TargetDiff} we measured the \textbf{Vina score}, \textbf{minimization}, and \textbf{dock} of each molecule, representing the energy from the generated structure, local energy minimization, and re-docking, respectively. Unlike Vina minimization, Vina dock performs a global search for the lowest-energy pose, losing dependency on the generated pose. Additionally, we report a \textbf{higher percentage}, indicating the ratio of generated molecules with lower Vina dock than the reference ligands.

\cref{tab:gen_reeval} shows that among the reference-free approaches, BInD outperformed the baselines in Vina score and Vina minimization, showcasing the generation of appropriate binding poses. 
For Vina dock and higher percentage metrics, BInD ranked within the top 2, closely matching the reference molecules. At the same time, BInD\textsuperscript{ref} showed the best results in most metrics, surpassing the other reference ligand-dependent approaches.
Moreover, while other models that receive spatial information from the reference ligand show significantly different values of binding affinities, BInD and BInD\textsuperscript{ref} maintained consistent results regardless of the reference ligand information, demonstrating their robustness as a distribution learner.

In particular, while BInD achieved Vina dock and its higher percentage performance on par with the baseline methods, it outperformed them in Vina score, with a value of -5.64 kcal mol$^{-1}$, and minimization energy, at -6.22 kcal mol$^{-1}$. As mentioned above, Vina dock loses dependence on the initially generated 3D pose, indicating that BInD’s generated molecular structures are less likely to be unstable poses that are highly positive in Vina score, as an example shown in the Figure \textcolor{\scolor}{S1}b. This observation prompted us to further analyze the energy differences between Vina scores and other Vina energies.

\cref{fig:gen_benchmark}a shows the distribution of energy differences between the Vina score and Vina dock. BInD and BInD\textsuperscript{ref} achieved the second-lowest and the lowest differences among the reference-free and reference-dependent methods, respectively. Since Vina dock energy relies solely on the molecular graph, these results suggest that BInD generates molecular structures well-aligned with their most stable poses, leading to the outperforming of other baselines. \cref{fig:gen_benchmark}b presents the energy shift following minimization from the generated pose, where BInD also shows the second-lowest energy change, indicating that its initial poses are in stable conformations within the protein pocket. Although AR demonstrated more minor energy differences in both cases, it may be due to the generation of rigid molecules with fewer rotatable bonds, as noted in Figure \textcolor{\scolor}{S6}. 

To further investigate the change in binding pose during energy minimization, we also report the cumulative density of RMSD between the generated and minimized poses. \cref{fig:gen_benchmark}c shows the results. Consistent with the energy difference trends of the Vina score and minimization, BInD achieved the second-best performance among the reference-free methods, highlighting the reliability of the generated poses attributed to our explicit consideration of interactions.\newline

\noindent\textbf{Molecular properties}. To assess the molecular properties of the generated molecules, we evaluated quantitative estimation of drug-likeness (\textbf{QED}) \cite{QED} and synthetic accessibility (\textbf{SA}) scores, \cite{SA} as well as the \textbf{diversity}. Since the graph structure solely determines molecular properties, we also measured the Fr\'echet ChemNet distance (\textbf{FCD}) \cite{ChemNetDistance} to measure the similarity between the generated and the training set molecules, validating models whether they have learned to generate drug-like molecules. Additionally, we examine the \textbf{functional group} composition of molecules.

\begin{table*}[!ht]
    \centering
    \scriptsize 
    
    \begin{adjustbox}{max width=\textwidth}
    \begin{tabular}{l|cc|cc|c|cc|cc}

        \toprule
        \multirow{2}{*}{Model} & \multicolumn{2}{c|}{QED ($\uparrow$)} & \multicolumn{2}{c|}{SA ($\uparrow$)} & \multirow{2}{*}{FCD ($\downarrow$)} & \multicolumn{2}{c|}{Func. Group ($\downarrow$)} & \multicolumn{2}{c}{Diversity} \\

        & Avg. & Med. & Avg. & Med. & & MAE & JSD & Avg. & Med. \\
        
        \midrule
        AR & 0.50 & 0.52 & 0.63 & 0.62 & 14.62 & 0.0564 & 0.2586 & 0.74 & 0.74 \\
        
        Pocket2Mol & \textbf{0.57} & \textbf{0.57} & \textbf{0.74} & \textbf{0.74} & 10.83 & 0.0383 &  0.2387 & 0.73 & 0.77 \\ 
        
        DiffSBDD & 0.49 & 0.50 & 0.62 & 0.62 & \textbf{10.77} & 0.0465 & 0.2362 & 0.79 & 0.79 \\
        
        TargetDiff & 0.49 & 0.49 & 0.61 & 0.59 & 18.38 & 0.0386 & 0.2692 & 0.74 & 0.73 \\ 
        
        DecompDiff & 0.49 & 0.49 & 0.61 & 0.59 & 11.27 & \textbf{0.0227} & \textbf{0.1821} & 0.74 & 0.73 \\
        
        \textbf{BInD} & \textbf{0.50} & \textbf{0.54} & \textbf{0.65} & \textbf{0.66} & \textbf{7.23} & \textbf{0.0311} & \textbf{0.1853} & 0.75 & 0.75  \\ 

        \midrule
        
        InterDiff & 0.32 & 0.30 & 0.56 & 0.57 & 17.36 & 0.0620 & 0.4626 & 0.75 & 0.75 \\ 
        
        DecompDiff\,\textsuperscript{ref} & 0.47 & 0.48 & 0.63 & 0.63 & 6.93 & 0.0329 & \textbf{0.1766} & 0.70 & 0.71 \\ 
        
        \textbf{BInD\textsuperscript{ref}} & \textbf{0.49} & \textbf{0.50} & \textbf{0.66} & \textbf{0.69} & \textbf{6.05} & \textbf{0.0302} & 0.1774 & 0.68 & 0.69 \\
        
        \midrule
        
        Reference & 0.48 & 0.47 & 0.73 & 0.74 & - & - & - & - & - \\
        
        \bottomrule
    \end{tabular}
    \end{adjustbox}
    \caption{
    \textbf{Benchmark result of BInD and baseline models on molecular properties.}
    The first six models are reference-free methods, and the last three are dependent methods. Mean and median values of QED, SA, FCD, functional group MAE and JSD, and diversity are reported. The top 2 performing models among reference-free methods and the best-performing model among reference-dependent models for each benchmark are marked as bold.}
    \label{tab:gen_topo}
\end{table*}

\cref{tab:gen_topo} presents the results where BInD achieved the second-best performance in both QED and SA scores among the reference ligand-free methods. While diffusion models generally fall behind auto-regressive models in molecular properties, BInD showed the best performance in generating drug-like molecules among diffusion-based baselines.

BInD achieved the best FCD score of 7.23, surpassing all baselines, and ranked second in functional group metrics (MAE: 0.0311, JSD: 0.1853). 
This can be attributed to BInD’s single-shot generation of 3D molecular graphs, allowing the designed molecules to closely align with the chemical distributions of the training set. 
While FCD and functional group distributions evaluate the model’s capacity to generate drug-like molecules, these distributions can also depend on the target pocket. We report FCD and functional group distributions for test set reference molecules in Supporting Information \textcolor{\scolor}{3.4} to address this. BInD outperformed all baselines, demonstrating its ability to generate molecules whose distribution closely resembles the distribution of the reference test molecules, implying its generalizability to unseen pockets. 
Comparing BInD with BInD\textsuperscript{ref} reveals the impact of providing a reference interaction pattern during the generation. While BInD generates diverse molecules, BInD\textsuperscript{ref} exhibited lower diversity but achieved better FCD and functional group matching compared to BInD, showing that providing reference interaction patterns can result in a more focused distribution of molecules. 

End-to-end 3D graph generative models show superior performance in molecular properties compared to the other baselines. By avoiding post-bond assignment tools, these models minimize potential biases. This is particularly important since different chemical bonds can have similar geometric properties but distinct contributions to molecular properties, such as CC double bond and aromatic CC bond. 
The results indicate that explicit bond generation helps learn the distribution of trained molecules, avoiding post-hoc bond assignment limitations that do not consider model-intended molecular properties. \newline

\noindent\textbf{Internal local geometry}. Previous studies \cite{TargetDiff, DecompDiff, Lingo3DMol} have mainly evaluated local structures in generated molecules through bond lengths, angles, and atom pairwise distances. We report these distributions in Figure \textcolor{\scolor}{S3-S5}, where BInD closely aligns with the referential local structure distribution. 

To evaluate the local structure more rigorously, we analyzed the strain energy of the generated molecules. 
\cref{fig:gen_benchmark}d presents the distribution of strain energies. Among the baseline models, BInD and BInD\textsuperscript{ref} achieved the second-lowest strain energy across all baselines. While diffusion-based models generally showed higher strain energies than auto-regressive models, BInD achieved the lowest among them. This result highlights the effectiveness of knowledge-based guidance terms, which capture the locality of molecular structure and refine atom positions in alignment with their bond linkages, leading to more stable molecular conformers. 

We further analyzed strain energies versus the number of rotatable bonds in Figure \textcolor{\scolor}{S7}. Auto-regressive models showed increased strain energy as more rotatable bonds grew, whereas bond-generating diffusion models (DecompDiff, BInD) maintained consistent strain energy levels. BInD matched Pocket2Mol when more than three rotatable bonds were present, demonstrating its robust performance in handling flexible molecular structures. \newline

\noindent\textbf{\revision{Success rate on three perspectives.}}
\revision{
In previous evaluations on molecular properties, interaction, and local geometry, BInD consistently demonstrated superior or competitive performance across all axes, underscoring the effectiveness of our multi-objective generation strategy.
To further quantify and evaluate the generated molecules that satisfy all criteria simultaneously, we report the success rate, using filters based on each criterion.
Starting from 10,000 generated molecules, and following previous studies \cite{DecompDiff,mars,desert}, we filtered molecules with QED $>$ 0.25, SA $>$ 0.59, and Vina docking score $<$ -8.18 kcal mol$^{-1}$ to satisfy the molecular property criteria.
Next, we excluded molecules with a difference between the Vina docking score and Vina minimized score exceeding 1.09 kcal mol$^{-1}$, indicating unstable or energetically unfavorable structures.
Finally, to assess local geometry, we removed molecules with strain energy exceeding 200 kcal mol$^{-1}$.
This multi-stage filtering reflects our focus on generating reliable molecules that satisfy all three key objectives within a single framework.
}

\revision{Table S1 presents the results. BInD achieved the highest success rate across all three criteria, with 4.7\% of generated molecules passing every filter, followed by Pocket2Mol (4.0\%). While Pocket2Mol outperformed diffusion-based baselines when only the molecular property filter was applied (25.6\%), its success rate dropped significantly when the interaction filter was added (6.7\%), which assesses the reliability of the generated 3D poses. In contrast, BInD retained a higher proportion of molecules after both interaction and local geometry filters, demonstrating its robustness in generating realistic and well-aligned molecular structures.}
\newline

\noindent\textbf{Generated examples}. \cref{fig:nci}e and Figure \textcolor{\scolor}{S2} show examples of generated molecules with higher binding affinities than that of the reference molecule. The molecules exhibit favorable molecular properties, surpassing the reference molecule in QED, SA, or both. 
In addition, they are diverse in structures, including complex ring systems, along with accurate 3D conformers. 
BInD demonstrates its strength in generating molecular structures that feature numerous NCIs, motivating us to further investigate the deep SBDD models in the point of NCIs at \cref{sec:NCI}.

\subsection{Elucidating the Role of NCIs in designing molecules}\label{sec:NCI}

\begin{figure}[!ht]
    \centering
    \includegraphics[width=0.9\linewidth]{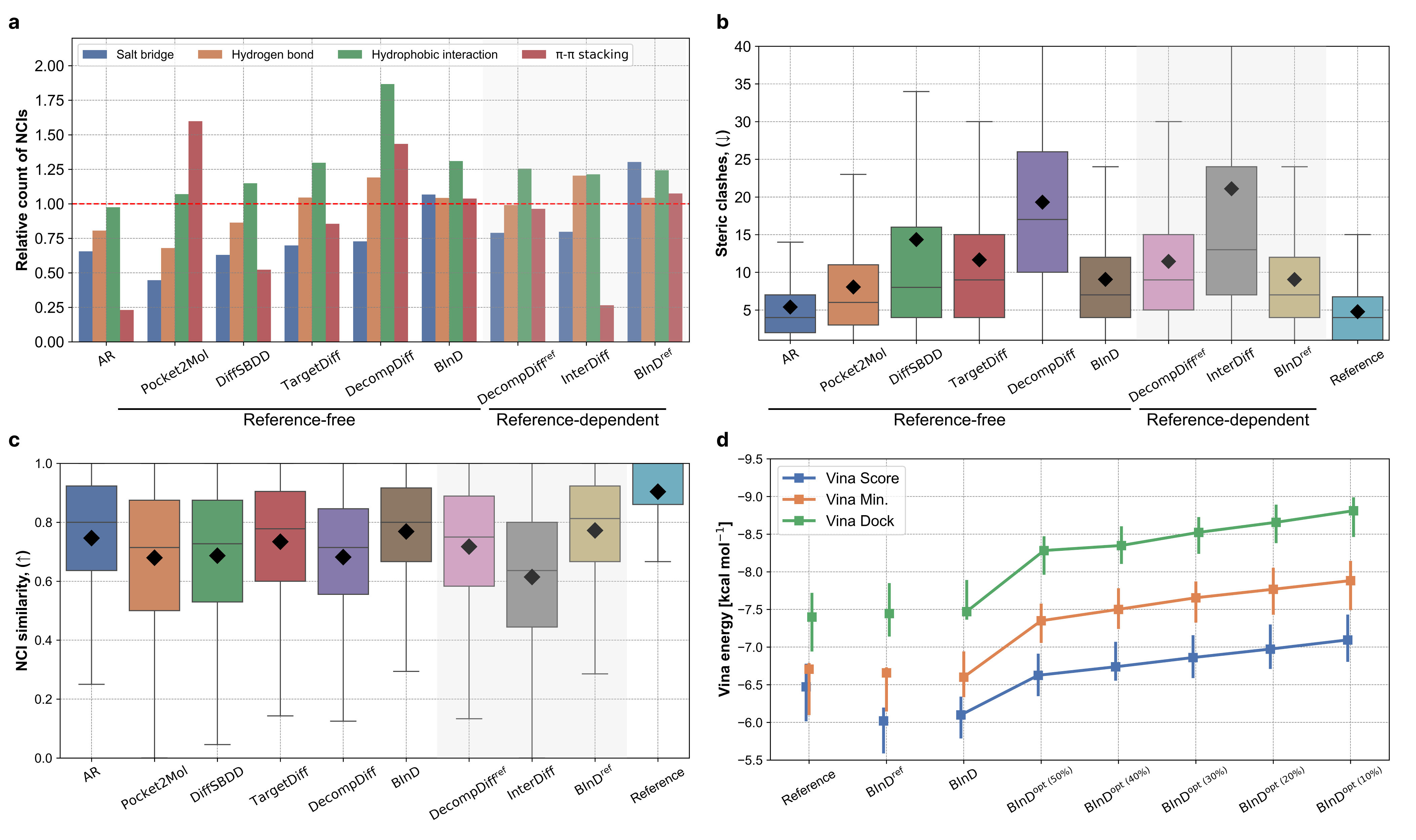}
    \caption{
    \textbf{Discerning the role of NCIs in SBDD.
    } 
    \textbf{a}, A histogram displaying the relative counts of NCIs averaged over the test data points. Each relative count is normalized to a value of 1.0 with the number of NCIs in the corresponding reference molecule. 
    \textbf{b}, A box plot of the number of steric clashes between the generated ligand and protein. 
    \textbf{c}, A box plot shows the NCI similarity of the generated and Vina minimized conformation. 
    For both box plots, median and four quartile values are described, with black diamonds indicating the average values. The reference-dependent models are differentiated in the shaded region.
    \textbf{d}, Vina score, minimization, and docking energies of reference molecules and molecules generated from the variants of BInD. As the ratio for NCI pattern retrieval from the initial generation decreases, the generated molecules consistently exhibit stronger bindings while preserving the gap between the three energy components.}
    \label{fig:nci} 
\end{figure}

A reliable SBDD demands a more detailed understanding of the molecular interface between a protein and a ligand, emphasizing the importance of NCIs.  
They enhance binding affinities and contribute to achieving stability and specificity to a target protein. \cite{nci_specific} For instance, the directionality of hydrogen bonds and salt bridges
can induce structural compatibility in the pocket alongside binding energies. 
In this regard, generating molecules with accurate binding modes is essential. Despite the significance of NCIs, a previous study \cite{PoseCheck} pointed out that deep SBDD models often struggle to generate molecules that both establish favorable NCIs and avoid steric clashes with protein atoms.\newline

\noindent\textbf{Evaluation of NCI}. We evaluate deep SBDD models on their abilities to generate favorable NCIs with fewer steric clashes. \cref{fig:nci}a and b present the results on the relative counts of NCIs and the number of steric clashes, respectively, with respect to the corresponding reference molecules. Consequently, achieving NCI counts as many as the reference molecule on each NCI type implies that a generated molecule can interact sufficiently with a pocket without neglecting potential sites.

Among all models, only BInD achieved relative NCI counts above 1.0 across all interaction types, while others showed deficiencies in more than one NCI type. AR, DiffSBDD, and TargetDiff performed significantly less well in salt bridges and $\pi$-$\pi$ stacking. 
Despite having the highest hydrophobic interaction count, DecompDiff showed more than twice as many steric clashes as BInD, indicating that its higher NCI count may come at the cost of structural instability.
All three generated similar NCI counts within the reference ligand-dependent models, except for $\pi$-$\pi$ stacking of InterDiff. 
DecompDiff\textsuperscript{ref} and BInD\textsuperscript{ref} showed comparable results, with BInD\textsuperscript{ref} showing fewer steric clashes, highlighting better structural reliability. Overall, BInD achieved sufficient balance across NCI types with low steric clashes.

Next, we evaluated the NCI similarity between the generated molecular structures and their energy-minimized poses. A reliable SBDD model should generate molecules with favorable NCIs to avoid structural changes during energy minimization. 
\cref{fig:nci}c illustrates the result. BInD showed the highest NCI similarity with a median of 0.8, demonstrating the effectiveness of explicitly considering NCIs during generation. 
\revision{Moreover, compared to InterDiff, BInD\textsuperscript{ref}, which is trained to simultaneously generate NCI patterns rather than simply using a given reference ligand's NCI pattern, demonstrates greater robustness in NCI-guided generation.}
\newline

\noindent\textbf{Molecular design empowered by NCIs}. 
While deep SBDD models generate molecules with desirable properties, as a distribution learner, they can fall short of achieving binding affinities higher than those of the training molecules. To address this, DecompOpt \cite{DecompOpt} suggested re-docking molecular fragments into protein pockets to maximize binding affinities, but this involves computationally intensive docking for each molecule. Drawing on real-world drug design scenarios, in which NCIs guide the design of high-affinity molecules, \cite{design_aminopyrimidinyl} we introduce a NCI-driven molecule design \revision{and optimization} approach that retrieves favorable NCI patterns from previously sampled molecules for the next sampling. The details of the NCI-retrieving strategy are provided in the Methods section. 

\cref{fig:nci}d displays Vina energies for molecules generated by BInD\textsuperscript{ref}, BInD, and BInD\textsuperscript{opt($p$\%)} with a series of $p$s from 50\% to 10\%. 
BInD showed higher Vina energies than those of reference molecules, while it achieved slightly lower Vina energies than those of BInD\textsuperscript{ref} by effectively identifying favorable NCI patterns. 
As $p$ decreases, retrieving NCI patterns with lower Vina scores, BInD\textsuperscript{opt($p$\%)} starts to produce molecules with higher affinities. The Vina score, minimization, and docking values maintain relative differences, showing that inpainting NCIs does not damage the performance of BInD. Notably, this NCI retrieval relied solely on in-place scoring without computationally intensive docking trials.\newline

\begin{figure}[!ht]
    \centering
    \includegraphics[width=0.94\linewidth]{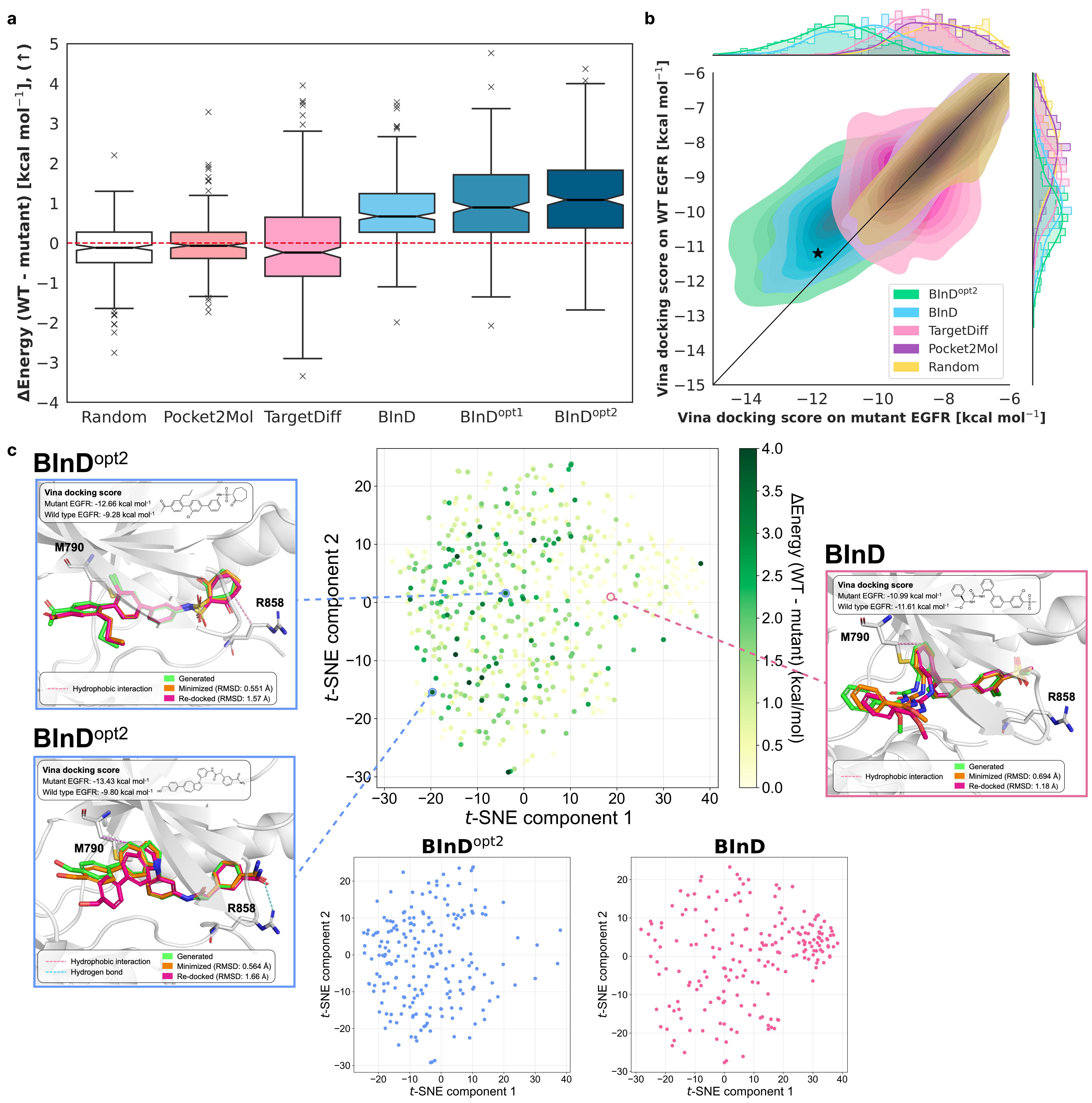}
    \caption{
    \textbf{BInD as a promising framework for designing mutant-selective EGFR inhibitors.} 
    \textbf{a}, A box plot illustrates differences between Vina docking scores on the mutant and WT EGFR pockets, where positive values correspond to the lower score on the mutant. Median values are depicted as notches as well as four quartile values. 
    \textbf{b}, Density plot showing the distributions of Vina docking scores on each mutant and WT EGFR pockets. A diagonal line divides a plot in half, where the upper area indicates where the docking score on the mutant pocket is lower. A black star indicates the reference ligand of its original crystal structure. 
    \textbf{c}, $t$-SNE plot in the middle visualizes the NCI patterns generated from BInD, BInD\textsuperscript{opt1}, and BInD\textsuperscript{opt2}, with the color indicating the selectivity measure. Two $t$-SNE plots at the bottom correspond to BInD (right, gray) and BInD\textsuperscript{opt2} (left, blue), respectively. Examples of generated molecules are visualized with Vina minimized and docking poses and Vina docking scores. NCIs with the mutated residues -- M790 and R858 -- are shown as dashed lines.
    }
    \label{fig:case_study}
\end{figure}

\noindent\textbf{Target-specific molecule design}. We applied our NCI-driven design strategy to a case that aims to selectively inhibit a \revision{double-}mutant epidermal growth factor receptor (EGFR) while sparing the wild-type (WT) receptor. Experimental details are provided in the Methods section.

\revision{As shown in \cref{fig:case_study}a, molecules generated by BInD showed marginally stronger binding to the mutant pocket with a median affinity difference of 0.67 kcal mol$^{-1}$. In contrast, the molecules generated by Pocket2Mol and TargetDiff showed -0.068 and -0.24 kcal mol$^{-1}$, respectively, exhibiting indistinguishable affinity differences compared to the randomly selected molecules from the training set, despite their target-aware generation.}

Optimizations further increased the binding energy difference, achieving median values of 0.89 and 1.1 kcal mol$^{-1}$ for BInD\textsuperscript{opt1} and BInD\textsuperscript{opt2}, respectively. 
\cref{fig:case_study}b \revision{apparently} shows that the binding energy distribution of BInD\textsuperscript{opt2} shifted toward \revision{mutant-selective region: upper left side of the diagonal line.} 
This \revision{reveals} that the selectivity between similar protein pockets can be effectively \revision{reinforced} by retrieving target-specific NCI patterns and \revision{undergoing optimization cycles.}
\revision{Moreover, the plot illustrates that the molecules generated from BInD readily possess stronger binding affinities toward the target than the other models. }
\revision{Additional result in Table S15 supports the claim, where the proportions of BInD-generated molecules in which the energy differences exceed a certain criterion increase with the optimization cycle. Note that only BInD\textsuperscript{opt1} and BInD\textsuperscript{opt2} can generate molecules with an energy difference exceeding 4 kcal mol$^{-1}$ among compared models.}
\revision{In contrast, the distribution of Pocket2Mol largely overlaps with that of random molecules, barely showing an advantage in utilizing the information of the target pocket.}

We visualized the NCI patterns of the generated molecules by plotting $t$-distributed stochastic neighbor embedding ($t$-SNE) of NCI fingerprints. The middle plot in \cref{fig:case_study}c shows the combined results for BInD, BInD\textsuperscript{opt1}, and BInD\textsuperscript{opt2}, with higher selectivity NCI patterns clustered on the left. Separate $t$-SNE plots for BInD and BInD\textsuperscript{opt2} below reveal that while BInD’s NCI patterns are evenly spread with a concentration on the right side, BInD\textsuperscript{opt2} shifts more to the left. This shift indicates that the NCI-driven sampling strategy increasingly focuses on the mutant-selective region through optimization.

We show examples of generated molecules on both sides of \cref{fig:case_study}c. In each example, Vina minimized and docked structures are shown together with RMSD values and Vina docking scores to the mutant and WT pockets. All of them showed well-aligned Vina minimization and docking poses, indicating the reliability of our model for generating stable poses. Two examples on the left side of \cref{fig:case_study}c are from BInD\textsuperscript{opt2}, where both consist of NCIs with two mutated residues, Met790 and Arg858, respectively. The example of BInD, located on the right side of \cref{fig:case_study}c, also involves a hydrophobic interaction with Met790. However, the molecule is too far from Arg858 to form an NCI. Upon thinking that the mutated residues are the main contributors that derive the structural difference between the mutant and WT pockets, forming the interaction with the residues may \revision{provide} a chance to gain selectivity, as in the example molecule from BInD. Intriguingly, \revision{these} NCIs are formed without expert knowledge of the mutation, which explains why BInD could achieve such high selectivity in contrast to the two baselines. Thus, our case study emphasizes the potential of BInD to design a target-specific molecule with emerging intelligence to detect critical sites that would lead to the desirable interaction.

\section{Conclusion}

In this work, we presented BInD, a diffusion-based framework for reliable SBDD that simultaneously addresses three critical objectives: local geometry, molecular properties, and protein-ligand interactions. By representing protein-ligand complexes as an informative bipartite graph that explicitly incorporates both bonds and interactions and formulating them into knowledge-based guidance terms, our approach achieves superior yet balanced performance across various benchmarks related to the three key objectives, highlighting the effectiveness of our strategy. BInD offers a well-rounded approach to molecular design, addressing both structural and interaction-related aspects of protein-ligand binding. The framework's ability to generate chemically feasible molecules with realistic 3D poses and favorable binding interactions opens new possibilities for computer-aided drug discovery and optimization. We envisage BInD as a promising example of leveraging the multi-task objectives to achieve a challenge to balance the performance between key requirements that are crucial in various practical scenarios. \cite{MultiObjectOptCatalyticCrack, BalancingEffTrans}

\section{Methods}

\noindent\textbf{BInD framework}. Initially, positions are sampled from an isotropic Gaussian distribution, and types are set to an auxiliary type. BInD then generates a 3D molecular graph ($L_t$) and its corresponding NCI pattern ($I_t$) at a time $t$ between T and 0. A single denoising step at a time $t$ is summarized in \cref{fig:overview}d. Given $L_t$, $I_t$, and a protein pocket $P$ as input, a dynamic interaction network denoises the input into the previous time step $t-1$. During the generation process, the dynamic interaction network captures information from the global semantics to local refinements by progressively decreasing the message-passing distance. From the obtained $L_{t-1}^\text{gen}$ and $I_{t-1}^\text{gen}$, we integrate chemical knowledge to attenuate the potential inconsistency between molecular geometry and its bonds and interactions as a form of guidance terms. Positions of atoms are slightly modified to prevent a malformed structure regarding bond and NCI types.\newline 

\noindent\textbf{Definitions and notations}. Our model constructs 3D molecular graphs and NCI types between ligand and protein atoms from a given protein structure. Specifically, a protein pocket is provided as $P = \{ (h_i^{\textnormal{P}}, x_i^{\textnormal{P}},e_{ij}^{\textnormal{P}}) \}_{i,j \in \{1,2,...,N_\text{P}\}}$ where $N_\text{P}$ denotes the number of protein atoms. $h_i^{\textnormal{P}}$ and $e_{ij}^{\textnormal{P}}$ denote a protein node and edge feature, respectively, and $x_i^{\textnormal{P}} \in \mathbb{R}^3$ denotes the position of a protein atom. The generated output, comprising a molecule and NCIs, is provided as $L = \{ (h_i^{\textnormal{L}}, x_i^{\textnormal{L}}, e_{ij}^{\textnormal{L}}) \}_{i, j \in \{1,...,N_\text{L}\}}$ and $I = \{\mathbf{i}_{ij}\}_{i \in \{1,...,N_\text{P}\}, j \in \{1,...,N_\text{L}\}} $ where $N_\text{L}$ denotes the number of ligand atoms, $h_i^{\textnormal{L}}$ and $e_{ij}^{\textnormal{L}}$ denotes a ligand node and edge feature, respectively, $x_i^{\textnormal{L}} \in \mathbb{R}^3$ denotes a position of a ligand atom, and $\mathbf{i}_{ij}$ is an NCI type between a protein and ligand atom. For simplicity, we abbreviate the notation of the generated ligand and its NCIs as $h, x, e$, and $\mathbf{i}$. The details of node and edge features used in this work are provided in Table \textcolor{\scolor}{S11} and \textcolor{\scolor}{S12}.\newline

\noindent\textbf{Forward diffusion process for atom, bond, and NCI}. Recognizing the importance of edge features (bonds and NCIs) in deep learning-based SBDDs, we incorporated a diffusion process for a complete bipartite graph of the protein-ligand complex. The NCI patterns of the protein-ligand complexes were first extracted using the Protein-Ligand Interaction Profiler (PLIP). \cite{PLIP} Four types of NCI are considered, including salt bridges, hydrogen bonds, hydrophobic interactions, and $\pi$-$\pi$ interactions. The salt bridges and hydrogen bonds exhibit directional properties, such as donor-acceptor or anion-cation pairs. We constructed a bipartite graph from these extracted NCIs and chemical bonds, with protein and ligand atoms as nodes and NCIs as edges.

Following the progress in diffusion models on a continuous space \cite{DDPM} and a categorical space, \cite{D3PM} at each diffusion time step $t$, an atom position is shifted towards the standard Gaussian distribution by adding small Gaussian noise. Categorical entities, including the atom type, bond type, and NCI types, are perturbed by adding small probabilities to a certain absorbing type. The forward diffusion process at a time step $t$ is defined as: 
    \begin{equation}
        \label{eq:forward_all}
        \begin{aligned}
            q(L_{t},I_{t} &\mid L_{t-1}, I_{t-1}, P) = q(h_{t} \vert h_{t-1}) \cdot q(x_{t} \vert x_{t-1}) \cdot q(e_{t} \vert e_{t-1}) \cdot q(\mathbf{i}_{t} \vert \mathbf{i}_{t-1})\text{,}
        \end{aligned}
    \end{equation}
while the forward diffusion process of continuous atom position $x_t$ and categorical $h_t$, $e_t$, $\mathbf{i}_t$ are denoted as follows:
    \begin{equation}
        \label{eq:cont_forward}
        q(x_t|x_{t-1}) = \mathcal{N}(x_t; \sqrt{1 - \beta_t} x_{t-1}, \beta_t \mathbf{{I}}),
    \end{equation}
    \begin{equation}
        \label{eq:cat_forward}
        q(\mathbf{i}_t|\mathbf{i}_{t-1}) = \mathcal{C}(\mathbf{i}_t; \mathbf{i}_{t-1} Q_t),
    \end{equation}
with the predefined noise schedule, $\beta_{1}, \cdots, \beta_{T}$, while $h_t$ and $e_t$ following the same equation. The probability transition matrix $Q_t$ introduces a small noise, shifting toward a certain probability mass, namely the absorbing type, by slightly decreasing the probability of $h_{t-1}$. For bonds and NCIs, the non-bonding and non-NCI types are naturally selected as the absorbing types, respectively, and for the atom types, an additional masking atom type, depicted as $U$ in \cref{fig:overview}c, is introduced as the absorbing type. With the $m$-th index as the absorbing type, the probability transition matrix is as follows:
    \begin{equation}
        \label{eq:Q_matrix}
        [Q_t]_{ij} =
        \begin{cases}
            1, & \text{if } i = j = m, \\
            1 - \beta_t, & \text{if } i = j \neq m, \\
            \beta_t, & \text{if } j = m, i \neq m.
        \end{cases}
    \end{equation}

The choice of the noise schedule for each of $[h, x, e, \mathbf{i}]$ varies in the implementation as described in Supporting Information \textcolor{\scolor}{8.1}; for convenience, we unified the notation $\beta_t$ regardless of the entity type. 
The protein-ligand complex's overall distribution is decomposed into node, position, edge, and NCI features, each treated as an independent distribution. By the Markov property, $L_{t}$ and $I_{t}$ can be directly obtained as a closed form equation from the data $L_0$ and $I_0$ with $\alpha_t = 1 - \beta_t$, $\overline{\alpha}_t = \prod_{s=1}^{t} \alpha_s$, and $\overline{Q}_t = Q_1Q_2 \ldots Q_t$ as follows:
    \begin{equation}
        q(x_t\vert x_0) = \mathcal{N}(x_t;\sqrt{\overline{\alpha}_t}x_0, (1 - \overline{\alpha}_t)\mathbf{{I}}),
        \label{eq:Markov_1}
    \end{equation}
    \begin{equation}
        q(\mathbf{i}_t\vert \mathbf{i}_0) = \mathcal{C}(\mathbf{i}_t;\mathbf{i}_0\overline{Q}_t),
        \label{eq:Markov_2}
    \end{equation}
where $e_t$ and $\mathbf{i}_t$ follow the same as \cref{eq:Markov_2}.
For a large $t$, the distribution of atom, bond types, and NCI types will be a point mass distribution to the absorbing type, while the distribution of atomic positions will become a standard Gaussian distribution.\newline

\noindent\textbf{Reverse generative process}. During the reverse generative process, a molecule is generated from the prior distribution, which is $\mathcal{N}(\mathbf{0}, \mathbf{I})$, centered on the protein pocket centroid for atom positions and a point mass categorical distribution on the absorbing type for categorical entities. Following previous works, \cite{DDPM, D3PM} a posterior of the single step of types and positions can be obtained by the Bayes rule as follows:
    \begin{equation}    
        q(x_{t-1}|x_{t},x_0) =\mathcal{N} (x_{t-1};\tilde{\mu}_t(x_t, x_0),\tilde{\beta}_t \mathbf{I}), 
        \label{eq:cont_posterior} 
    \end{equation}
    \begin{equation}        
        q(\mathbf{i}_{t-1}|\mathbf{i}_{t},\mathbf{i}_0) = \mathcal{C}\biggl(\mathbf{i}_{t-1} ; \frac{\mathbf{i}_{t}Q_{t}^{\top} \odot \mathbf{i}_{0}\overline{Q}_{t-1}}{\mathbf{i}_{0} \overline{Q}_t \mathbf{i}_{t}^\top }\biggr), \label{eq:cat_posterior}
    \end{equation}
where
    \begin{equation}
        \tilde{\mu}_t(x_t, x_0) := \frac{\sqrt{\overline{\alpha}_{t-1}} \beta_t}{1 - \overline{\alpha}_t} x_0 + \frac{\sqrt{\alpha}_t (1 - \overline{\alpha}_{t-1})}{1 - \overline{\alpha}_t} x_t, \label{eq:mu}
    \end{equation}
    \begin{equation}
        \tilde{\beta}_t := \frac{1 - \overline{\alpha}_{t-1}}{1 - \overline{\alpha}_t} \beta_t, \label{eq:beta_tilde}
    \end{equation}
while the posterior for bond and atom types follows the same as \cref{eq:cat_posterior}. To match the true reverse generative posterior, we parameterize the reverse process, $p_\theta(L_{t-1},I_{t-1}|L_{t},I_{t}, P)$. Multiple choices can parametrize the reverse process; we chose the model to predict the original data $[L_0, I_0]$ following the previous diffusion-based molecular generative models. \cite{MolDiff, TargetDiff}
Specifically, starting from $L_0$ and $I_0$, we first sample $L_t$ and $I_t$ by \cref{eq:Markov_1,eq:Markov_2}. The model $\mathcal{X}_\theta$ predicts $\hat{L}_0$ and $\hat{I}_0$, with the training objective defined as follows:
    \begin{equation}
        \hat{L}_0, \hat{I}_0 = \mathcal{X}_{\theta}(L_{t},I_{t},t,P),
    \end{equation}  
    \begin{equation}
        \text{Loss} = D_\text{KL}\bigl(q(L_{t-1} | L_t, L_0) \Vert p_\theta(L_{t-1} | L_{t}, \hat{L}_{0})\bigr) 
        + D_\text{KL}\bigl(q(I_{t-1} | I_t, I_0) \Vert p_\theta(I_{t-1} | I_{t}, \hat{I}_{0})\bigr).
    \end{equation}
The hyper-parameter setting for training is detailed in Table \textcolor{\scolor}{S13}.\newline

\noindent\textbf{Dynamic interaction network}. Strongly inspired by the recent model architectures \cite{EGNN, SchNet, TFN} that account for a molecule with 3D coordinates, we design a new $E(3)$-equivariant neural network to denoise a 3D molecular graph inside a protein pocket. While EGNN \cite{EGNN} originally updates the hidden features of node and atom positions in a homogeneous graph, protein-ligand complexes are represented as bipartite graphs with heterogeneous NCI edges. Moreover, the number of edges for NCI predictions scales to $N_\text{P} \times N_\text{L}$, though the actual NCIs to predict are sparse. As a result, it is essential to design an $E(3)$-equivariant model architecture that efficiently propagates messages through the bipartite graph. 

To accomplish this, for a noised graph as input, we first construct a radial graph with distinct distance cutoffs for protein-protein nodes, protein-ligand nodes, and ligand-ligand nodes, denoted as $\gamma_{\text{P}}$, $\gamma_{\text{I}}$, and $\gamma_{\text{L}}$, respectively: 
    \begin{equation}
        m_{ij}^{\text{P}} \leftarrow f_{\phi}^{\text{P}} (h_i^\text{P}, h_j^\text{P}, e_{ij}^{\text{P}}, d_{ij}^{\text{P}}),
    \end{equation}

    \begin{equation}
        m_{ij}^{\text{L}} 
        \leftarrow f_{\phi}^{\text{L}} (h_i^\text{L}, h_j^\text{L}, e_{ij}^{\text{L}}, d_{ij}^{\text{L}}), 
    \end{equation}

    \begin{equation}
        m_{ij}^{\text{I}} \leftarrow f_{\phi}^{\text{I}} (h_i^\text{P}, h_j^\text{L}, \mathbf{i}_{ij}, d_{ij}^{\text{I}}). 
    \end{equation}
The message features are then separately aggregated to each protein and ligand node as follows:
    \begin{equation}
        \label{eq:message_aggregate}
        m_i^{\text{P}}\leftarrow\sum_{j\in n_{\gamma_\text{P}}(i)}{m_{ij}^{\text{P}}}, 
    \end{equation}
    \begin{equation}
        m_i^{\text{L}}\leftarrow\sum_{j\in n_{\gamma_\text{L}}(i)}{m_{ij}^{\text{L}}},
    \end{equation}
    \begin{equation}
        m_i^{\text{I}},m_j^{\text{I}}\leftarrow\sum_{j\in n_{\gamma_\text{I}}(i)}{m_{ij}^{\text{I}}}, \sum_{i\in n_{\gamma_\text{I}}(j)}{m_{ij}^{\text{I}}}, 
    \end{equation}       
where $n_{\gamma}(i)$ is a set of atom indices, $\{j\}$, within the distance $d_{ij} < \gamma$. Note that the aggregation of $m_{ij}^{\text{I}}$ is asymmetric since an NCI edge is heterogeneous. For clarity, we use $i$ as an index for a protein atom and $j$ as an index for a ligand atom. Then, hidden node features are updated as follows:
    \begin{equation}
        h_i^\text{P} \leftarrow f_{\psi}^{\text{P}} (h_i^\text{P},  m_{i}^{\text{P}},  m_{i}^{\text{I}}, z^\text{P}, z^\text{L}),
    \end{equation}
    \begin{equation}
        h_j^\text{L} \leftarrow f_{\psi}^{\text{L}} (h_j^\text{L},  m_{j}^{\text{L}}, m_{j}^{\text{I}}, z^{\text{P}}, z^{\text{L}})\text{,} 
    \end{equation}
where the global protein and ligand features ($z^{\text{P}}$ and $z^{\text{L}}$) are computed by aggregating all the node features of protein and ligand nodes, followed by a shallow linear layer:
    \begin{equation}
        z^{\text{P}}, z^{\text{L}} = f_{\theta}\left(\sum {h_j^{\text{P}}}\right), f_{\theta}\left(\sum {h_i^{\text{L}}}\right)\text{.}
    \end{equation}
In the early stages of our experiments, we found that incorporating the compressed representation of protein and ligand features enhanced the training process, as the radial graph alone may not capture the global states of the protein and ligand. The ligand node positions and edges are then updated as follows:
    \begin{equation}
        x_j^\text{L} \leftarrow x_j^\text{L} + \sum_{i \in n_t(j)} f_{\tau}^{ \text{I}}(m_{ij}^{\text{I}}) \cdot (x_i^{\text{P}} - x_j^\text{L}) + \sum_{k \in n_t(j)}   f_{\tau}^{\text{L}}(m_{jk}^\text{L}) \cdot (x_k^\text{L} - x_j^\text{L}),
    \end{equation}
    \begin{equation}
        e_{ij}^\text{P}\leftarrow e_{ij}^\text{P} +  f_{\eta}^{\text{P}}(m_{ij}^\text{P}), 
    \end{equation}
    \begin{equation}
        e_{ij}^\text{L}\leftarrow e_{ij}^\text{L}+ f_{\eta}^{\text{L}}(m_{ij}^\text{L}),
    \end{equation}
    \begin{equation}
        \mathbf{i}_{ij}\leftarrow \mathbf{i}_{ij} + f_{\eta}^{\text{I}}(m_{ij}^{\text{I}}).
    \end{equation}    
Note that the subscript $\theta, \phi$, $\tau$, and $\eta$ indicate learnable model parameters.

During the reverse diffusion process, early steps (large $t$) are known to generate the global semantics of the sample, while later steps (small $t$) focus on fine-tuning and refinement. \cite{task_oriented_diff} To account for this, we set the radius cutoffs of \revision{ligand bond and interaction edges}
as increasing functions of the time step $t$. This dynamic cutoff allows for long-range message passing during the early steps (large $t$) to capture the global semantics of the protein, while later steps (small $t$) emphasize the local structure. We choose the simplest linear function:
    \begin{equation}
    \gamma(t) = \gamma_{\text{min}} + \frac{t}{T}(\gamma_{\text{max}} - \gamma_{\text{min}}).
    \label{eq:dynamic_cut}
    \end{equation}
The values for $\gamma_{min}$ and $\gamma_{max}$ are set to different values for bond and interaction edges provided in Table \textcolor{\scolor}{S13}. 
\revision{We further conduct an ablation study to differentiate the effect of the time-scheduled radius cutoffs. As variants of BInD, we train and sample using both maximum and minimum fixed distance cutoffs. The results in Table S9 demonstrate that the dynamic cutoff is beneficial in balancing the global interaction and local geometry properties of generated molecules.}

After processing through multiple layers of dynamic interaction networks, the types of atoms, bonds, and NCIs are predicted as follows:
    \begin{equation}
        h_{i, \text{one-hot}}^{\text{L}} \sim \mathcal{C} \left( \text{softmax}(\text{MLP}({h}_i^{\text{L}}))\right),
    \end{equation}
    \begin{equation}
        e_{ij,\text{one-hot}}^{\text{L}} \sim \mathcal{C} \left( \text{softmax}(\text{MLP}(e_{ij}^{\text{L}} + e_{ji}^{\text{L}}))\right), 
    \end{equation}
    \begin{equation}
        \mathbf{i}_{ij,\text{one-hot}} \sim \mathcal{C} \left( \text{softmax}(\text{MLP}(\mathbf{i}_{ij}))\right).    
    \end{equation}
\revision{The number of used dynamic interaction network layers is provided in Table S13.}\newline
    
\noindent\textbf{Sampling techniques}. After training the reverse generative model, we applied additional techniques to generate a large bipartite graph. When simultaneously generating both the molecular graph and its 3D structure, inconsistencies may arise between ligand atom-bond relationships and protein-ligand atom interactions, including NCIs. To address this issue, we employed two approaches: resampling and the use of guidance terms.

Since the SBDD tasks include both local (e.g., bond distance) and global (e.g., QED and SA) objectives, we incorporate repetitive resampling of forward and reverse steps during the sampling process. Previous work \cite{RePaint} demonstrated that, in the context of diffusion models for fixed image inpainting, iteratively refining intermediate states allows the model to better align global information with its local information. Building on this idea, we observed that resampling the intermediate states in a straightforward generation scheme significantly reduces inconsistencies between the generated bonds and atom positions, improving structural coherence. However, resampling throughout the reverse process would slow down the generative process due to the additional refinement iterations. Since the early stages of the reverse process already have sufficient steps for refinement, we opted to apply the resampling technique only during the later stages, where the fine-level generation occurs. 

Previous studies in diffusion-based image generation \cite{diffusion_beat_gans, GLIDE_image} have demonstrated that using guidance terms derived from the derivative of a classifier can substantially enhance sample quality. Building on this concept, MolDiff \cite{MolDiff} demonstrated that incorporating guidance terms based on bond distance deviations from predefined lengths improves unconditional molecular generation. Expanding on this idea, we extend the guidance terms to account for bond distances, bond angles, NCIs, and protein-ligand clashes. Specifically, we propose four \revision{knowledge-based} guidance terms: bond distance, bond angle, NCI distance, and steric clash guidance, defined as follows:
    \begin{equation}
        \mu_\theta'(x_t,t)=\mu_\theta(x_t,t)+\delta_{\text{BD}} + \delta_{\text{ID}} + \delta_{\text{BA}} + \delta_{\text{SC}}, 
    \end{equation}
    \begin{equation}
        x_{t-1}\sim\mathcal{N}(x_{t-1};\mu_\theta'(x_t,t),\tilde{\beta}_t \mathbf{I}),
    \end{equation}
where $\mu_\theta(x_t,t)$ is a model-predicted $\tilde{\mu}_t(x_t,x_0)$, while $\delta_{\text{BD}}$, $\delta_{\text{ID}}$, $\delta_{\text{BA}}$, and $\delta_{\text{SC}}$ denote the knowledge-based guidance terms.
Specifically, each guidance term is defined as follows:
    \begin{equation}
        \delta_{\text{BD}, i} = -a_{\text{BD}} \nabla_{x_t} \sum_{j \in n_{\hat{L}_0}(i)} \biggl[ \max(0, d_{ij}^{\text{L}} - d_\text{BD}^\text{max}) + \max(0, d_\text{BD}^\text{min} - d_{ij}^{\text{L}})\biggr], 
    \end{equation}
    \begin{equation}
        \delta_{\text{ID}, i} = -a_{\text{ID}} \nabla_{x_t} \sum_{j \in n_{\hat{L}_0}(i)} \sum_{\kappa \in \text{NCI}} \biggl[\max(0, d_{ij}^{\text{I}} - d_{\text{ID},\kappa}^\text{max}) + \max (0, d_{\text{ID},\kappa}^\text{min} - d_{ij}^{\text{I}})\biggr],
    \end{equation}
    \begin{equation}
        \delta_{\text{BA}, i} = -a_{\text{BA}} \nabla_{x_t} \sum_{\substack{i \in n_{\hat{L}_0}(j), \\ k \in n_{\hat{L}_0}(j), \\ i \neq k}}  \max(0, d_{\text{BA}}^{\text{min}} - d_{ik}^{\text{L}}), 
    \end{equation}
    \begin{equation}
        \delta_{\text{SC}, i} = -a_{\text{SC}} \nabla_{x_t} \sum_{i\in\{1,...,N_L\}} \sum_{j\in\{1,...,N_P\}} \max(0, d_\text{SC}^\text{min} - d_{ij}),
    \end{equation}
where $n_{\hat{L}_0}(i)$ denotes the indices of bond-connected nodes to $i$-th node in $\hat{L}_0$. $a_{\text{BD}}$, $a_{\text{ID}}$, $a_{\text{BA}}$, and $a_\text{SC}$ are scaling coefficients for each guidance term whose values are provided in Table \textcolor{\scolor}{S13}. Distance threshold values, $d_\text{BD}^\text{min}$, $d_\text{BD}^\text{max}$, $d_{\text{ID},\kappa}^\text{min}$,
$d_{\text{ID},\kappa}^\text{max}$,
$d_\text{BA}^\text{min}$,
and $d_\text{SC}^\text{min}$, are provided in Table \textcolor{\scolor}{S14}.
\revision{We conduct ablation studies on each term to clarify the contribution. We categorized guidance terms into intra- and inter-molecular geometry guidance, and their impact is evaluated by measuring performance changes on related benchmark metrics when those terms are neglected. Table S7 and S8 show the results, where intramolecular guidance terms are beneficial to lower the strain energy, while intermolecular guidance terms contribute to lowering the Vina score and steric clashes.}
\newline

\noindent\textbf{NCI-driven molecule design \revision{and optimization}}. In traditional molecule design, molecules are often created by anchoring or forbidding specific NCIs. \cite{design_aminopyrimidinyl} Previous methods have adopted this approach by conditioning molecule generation within a protein pocket based on the information of specific NCIs. \cite{DeepICL, InterDiff, Lingo3DMol} However, BInD simultaneously generates NCI edges along with a molecule, enabling the adoption of the inpainting technique. \cite{RenderDiffusion, RePaint} In light of this, we propose an NCI-driven molecule generation approach with desired NCI patterns.

The lower branch of \cref{fig:overview}d illustrates the case where a desired pattern of NCIs, $I_0^\text{fix}$, is given. The information is first noised to match the time step, then mixed with $I_{t-1}^\text{gen}$ with a binary mask matrix, $m^\text{fix}$, that indicates which interaction edge to fix with a given NCI. In summary, a single step of NCI inpainting can be formulated as follows:
    \begin{align}
        I^{\text{fix}}_{t-1} &\sim q(I^{\text{fix}}_{t-1} | I_t, I^{\text{fix}}_{0}), \label{eq37} \\
        I_{t-1}^{\text{gen}} &\sim q(I_{t-1}^{\text{gen}} | I_t, \hat{I}_{0}), \label{eq38} \\
        I_{t-1} &= m^{\text{fix}} \odot I^{\text{fix}}_{t-1} + (1 - m^{\text{fix}}) \odot I_{t-1}^{\text{gen}}. \label{eq39}
    \end{align}    
When a reference ligand is given, as in the test set, we extract its NCI pattern to set as $I_0^\text{fix}$. By letting all entities of $m^\text{fix}$ as 1, we can guide the molecular generation to follow the NCI pattern from the reference. We named the method as BInD\textsuperscript{ref}.

Moreover, we propose an NCI-retrieving strategy \revision{for optimizing target specificity}. 
This strategy begins by generating 100 molecules for each protein pocket. We then perform \revision{computationally cheap} in-place Vina scoring for all generated ligands, which is much faster than docking. We filter the top p \% of molecules with the highest affinities based on Vina scores and retrieve their NCI patterns from this. We randomly selected one NCI pattern for each generation round. This NCI pattern is set as $I_0^\text{fix}$ to drive a molecular generation process to increase binding affinity. Then, we re-generate molecules by following \cref{eq37,eq38,eq39}. For each $p$, we named the corresponding method BInD\textsuperscript{opt(p \%)}. \revision{In comparison to optimization strategies that retrieve molecular fragments for the next round of generation, \cite{DecompOpt} our method can search for more diverse molecules while retaining favorable interaction patterns, reducing the risk of mode collapse.}

Even though we here utilized whole NCI patterns from reference or pre-sampled molecules, one can also employ specialized knowledge on the target system to partially fix and generate the rest of the NCI patterns by setting $m^\text{fix}$, purposing to anchor or to veto specific amino acid residues.\newline

\noindent\textbf{Dataset}. We used the CrossDocked2020 dataset \cite{CrossDocked} to train and test BInD. Following previous works, \cite{AR, TargetDiff}, we curated a \revision{full CrossDocked2020} set of 22.5 million docked protein-ligand complexes, selecting those with a binding pose RMSD of less than 1 \si{\angstrom}. 
\revision{Then, we split the train and test sets with sequence identity below 30\%, resulting in a training set of 100,000 complexes comprising 14,365 protein structures of 1,419 gene families, and a test set of 100 different proteins and 91 gene families.}\newline

\noindent\textbf{Baselines}. We selected a range of deep SBDD models as baselines, including both autoregressive and diffusion models, as well as point cloud-generating and 3D graph-generating approaches. For autoregressive models, we choose AR, \cite{AR} which generates molecules by sequentially creating point clouds, and Pocket2Mol, \cite{Pocket2Mol} which generates a 3D graph.

For diffusion-based models using point cloud representations, we selected DiffSBDD \cite{DiffSBDD} and TargetDiff, \cite{TargetDiff} both of which generate atom types and positions within a protein pocket. DiffSBDD uses continuous diffusion, \cite{DDPM} while TargetDiff employs categorical diffusion \cite{D3PM} for atom types. DecompDiff \cite{DecompDiff} generates a molecule starting from decomposed priors to generate fragment graphs from each, which are then connected to form a complete molecule. DecompDiff offers two sampling options: one using fragment centers obtained from a prior sampling method \cite{AlphaSpace} (DecompDiff), and the other using fragment centers retrieved from reference ligands of the test complexes in the CrossDocked2020 dataset (DecompDiff\textsuperscript{ref}). InterDiff \cite{InterDiff} extends the point cloud-generating approach by generating molecules conditioned on an interaction prompt retrieved from the reference ligands' NCI patterns during generation.
\revision{Although InterDiff incorporates NCIs while generation, two important originality exist in our model: first, BInD generates an NCI pattern so that it is free from the reference, and second, geometric properties of each NCI type are considered during the generation.}

For all baseline models, we brought the models from their official repositories and generated molecules following the provided default settings. For a fair comparison, we separately compare methods requiring a reference ligand for a generation, including InterDiff, DecompDiff\textsuperscript{ref}, with BInD\textsuperscript{ref}, in which an NCI pattern from a reference ligand is inpainted.\newline

\noindent\textbf{Evaluation Details}. \revision{Following previous SBDD methods \cite{DiffSBDD,AR,DecompDiff,Pocket2Mol,TargetDiff}}, we \revision{generate} 100 molecules per \revision{100} CrossDocked2020 test pockets to evaluate BInD and baseline models across all metrics. Protein-ligand \revision{binding affinities} were evaluated using AutoDock Vina. \cite{AutoDockVina} 
\revision{Despite the existing uncertainties in affinity estimation of docking methods, the CrossDocked2020 dataset is intrinsically constructed via docking protocols, and a sufficient number of samples can make it statistically meaningful, rendering the assessment feasible.}
For the benchmark of molecular properties, QED and SA scores were calculated using RDKit, following the \revision{previous studies \cite{TargetDiff,Pocket2Mol}}. For each protein pocket, we computed the average Vina score, minimization, docking energy, QED, and SA values, and reported the overall average and median values. 

We further examined FCD and a functional group distribution to assess the chemical similarity between the generated and training data. Since the CrossDocked2020 dataset comprises drug-like molecules, \cite{CrossDocked} a lower FCD and well-aligned functional group distribution suggest that the generated molecules closely resemble drug-like compounds. The functional group distribution was assessed using JSD and MAE, following recent work. \cite{D3FG} Specifically, we selected the 25 most frequent functional groups in CrossDocked2020. Then, we compared two types of functional group distributions: frequency and ratio. The frequency distribution, evaluated with JSD, captures the overall occurrence pattern of each functional group, while the ratio, assessed with MAE, measures the average occurrence per molecule.

To evaluate strain energy, we followed the recent approach, \cite{PoseCheck} calculating it as $E_{\text{strain}} = E_{\text{generated}} - E_{\text{minimized}}$. Here, $E_{\text{generated}}$  is the universal force field (UFF) energy of the generated structure, and $E_{\text{minimized}}$  is the UFF energy after minimization, assessed in the absence of the protein. This evaluation provides insight into the model’s ability to produce stable and realistic local geometries. NCIs were analyzed using the protein-ligand interaction profiler (PLIP), \cite{PLIP} and steric clashes were evaluated following the PoseCheck \cite{PoseCheck} protocol. To calculate NCI similarity, we flattened the NCI patterns of the protein side into a 1D vector following the previous work \cite{DeepICL} and measured the Jaccard similarity between the vectors.\newline

\noindent\textbf{Details on \revision{double-}mutant EGFR\revision{-targeted} design}. We used two EGFR structures from Sogabe et al., \cite{EGFR} one is a wild-type (PDB ID: 3W2S), and the other is a double-mutant (PDB ID: 3W2R) where Thr790 is modified to a methionine and Leu858 is modified to an arginine. This double-mutated EGFR is clinically significant due to its role in exhibiting drug resistance. \cite{EGFR2} 
Initially, \revision{we generate 300 molecules} inside the pocket of mutant EGFR with BInD and the other two baseline models, Pocket2Mol and TargetDiff.
\revision{Although EGFR structures are included in the training set, all models -- BInD, Pocket2Mol, and TargetDiff -- were trained on the same dataset, ensuring a fair and consistent basis for comparison.}
Then, following the NCI-driven designing scheme that retrieves interaction patterns of the top 10 \% Vina scores, we carried out two rounds of optimization. At each round, 300 molecules are additionally generated, and each set of molecules is named BInD\textsuperscript{opt1} and BInD\textsuperscript{opt2}, where the numbers indicate the optimization round. 
\revision{In addition, as a negative control, we randomly select 300 compounds from the training set ligands so that they are intrinsically unrelated to the target.}
To assess selectivity, we docked molecules to both mutant and WT pockets. Both pockets share a reference ligand, so they have nearly identical pockets except for two mutated residues. \cite{EGFR} We \revision{use docking instead of} local minimization to prevent bias toward a pocket where molecules are generated.

\medskip
\textbf{Supporting Information} \par 
Supporting Information is available from the Wiley Online Library or from the author.

\medskip
\textbf{Acknowledgements} \par 

This work was supported by the Basic Science Research Program through the National Research Foundation of Korea, which is funded by the Ministry of Science and ICT (Grant No. NRF-2022M3J6A1063021; RS-2023-NR077040; RS-2023-00257479), and by the Ministry of Health and Welfare (Grant No. RS-2024-00512498), to J.L., W.Z., J.S., and W.Y.K.

\medskip
\textbf{Data and Code Availability} \par 

The data and processing scripts for BInD are available at \href{https://github.com/lee-jwon/BInD}{https://github.com/lee-jwon/BInD}. The original CrossDocked2020 dataset can be downloaded from \href{https://github.com/gnina/gnina}{https://github.com/gnina/gnina}.
The training and generation scripts for BInD and its trained settings are available at \href{https://github.com/lee-jwon/BInD}{https://github.com/lee-jwon/BInD}.

\medskip
\textbf{Author contributions} \par

J.L. and W.Z. conceptualized the work and developed the model. J.L. and W.Z. trained the model and carried out the experiments for benchmark and case study. J.L., W.Z., J.S., and W.Y.K. designed the experiments, analyzed the results, and contributed to manuscript writing. The whole work was supervised by W.Y.K.

\medskip

\bibliographystyle{MSP}
\bibliography{main}

\end{document}